 \documentclass[final,3p,times,twocolumn]{elsarticle}
   \usepackage{graphicx}
\usepackage{amssymb}
\journal{Physics Letter B}

\begin{document}

\begin{frontmatter}



\title{A convenient implementation of the overlap between arbitrary Hartree-Fock-Bogoliubov vacua for projection}


\author[1]{Zao-Chun Gao\corref{cor1}}
\ead{zcgao@ciae.ac.cn}
\cortext[cor1]{Corresponding author}
\author[1] {Qing-Li Hu}
\author[1] {Y. S. Chen}

\address[1] {China Institute of Atomic Energy, P.O. Box 275 (10), Beijing 102413, P.R. China}

\begin{abstract}
Overlap between Hartree-Fock-Bogoliubov(HFB) vacua is very important
in the beyond mean-field calculations. However, in the HFB
transformation, the $U,V$ matrices are sometimes singular due to the
 exact emptiness ($v_i=0$) or full occupation ($u_i=0$) of some single-particle orbits.
 This singularity may cause some problem in evaluating the overlap between
 HFB vacua through Pfaffian.
We found that this problem can be well avoided by setting those zero
occupation numbers $u_i,v_i$ to some tiny values denoted by
$\varepsilon(>0)$, which numerically satisfies $1+\varepsilon^2=1$
(e.g., $\varepsilon=10^{-8}$ when using the double precision data
type). This treatment does not change the HFB vacuum state because
$u_i^2,v_i^2=\varepsilon^2$ are numerically zero relative to 1.
Therefore, for arbitrary HFB transformation, we say that the $U,V$
matrices can always be nonsingular. From this standpoint, we present
a new convenient Pfaffian formula for the overlap between arbitrary
HFB vacua, which is especially suitable for symmetry restoration.
Testing calculations have been performed for this new formula. It
turns out that our method is reliable and accurate in evaluating the
overlap between arbitrary HFB vacua.
\end{abstract}


\end{frontmatter}

% \linenumbers


\section{Introduction}
\label{s1}
The Hartree-Fock-Bogoliubov (HFB) approximation has been
 a great success in understanding interacting many-body quantum
systems in all fields of physics. However, the beyond mean-field
effects (e.g., the nuclear vibration and rotation) are missing in
the HFB calculations. Methods that go beyond mean-field,
 such as the Generator Coordinate Method(GCM) and the projection method,
are expected to take those missing effects into consideration and
present better description of the many-body quantum system. In the
beyond mean-field calculations, operator matrix elements and
overlaps between multi-quasiparticle HFB states are basic blocks.
These matrix elements and overlaps can be evaluated using the
generalized Wick's theorem (GWT)\cite{Balian69,Hara79}, or
equivalently using Pfaffian
\cite{Oi12,Mizusaki12,Mizusaki13,Avez12,Bertsch12}, or using the
compact formula in Ref.\cite{Perez07}. However, in the efficient
calculations (e.g., see \cite{Mizusaki13}), all of the matrix
elements and overlaps require the value of the overlap between HFB
vacua.

Thus, the reliable and accurate evaluation of the overlap between
HFB vacua is very important for the stability and the efficiency of
the beyond mean-field calculations. Especially in cases near to the
Egido pole \cite{Egido01}, the overlap between HFB vacua is very
tiny, and a small error could lead to a large uncertainty of the
matrix elements. In the past, numerical calculations of the overlap
were performed with the Onishi formula \cite{Onishi66}.
Unfortunately, the Onishi formula leaves the sign of the overlap
undefined due to the square root of a determinant. Several efforts
have been made to overcome this sign problem
\cite{Neerg88,Ha92,Do98,Be08,Ha82,Oi05}. In 2009, Robledo proposed a
different overlap formula with the Pfaffian rather than the
determinant \cite{Robledo09}. This formula completely solves the
sign problem
 but requires the inversion of the matrix $U$ in the Bogoliubov transformation.
To avoid the singularity of $U$, the formula for the limit when
several orbits are fully occupied is given in Ref.
\cite{Robledo11}. Simultaneously, the limit when some orbits are
exact empty was also considered to reduce the computational cost.
Meanwhile, various Pfaffian formulae for the overlap between HFB
vacua have been proposed by several authors
\cite{Oi12,Avez12,Bertsch12}. In Ref. \cite{Bertsch12}, the overlap
formula does not require the inversion of $U$, but the empty orbits
in the Fock space should be omitted.

In practical calculation, one should first identify the singularity of the matrices $U$ and $V$
 in the Bogoliubov transformation.
 This can be easily tested with the Bloch-Messiah theorem (see details in Ref. \cite{Ring80}).
The matrices $U$ and $V$ can be decomposed as $U=D\bar U C$ and
$V=D^*\bar V C$. Here, $D$ and $C$ are unitary matrices. $\bar U$
and  $\bar V$ refer to the BCS-transformation and are constructed
from the occupation numbers $u_i, v_i$  with $0\leq u_i, v_i\leq1$
and $u_i^2+v_i^2=1$ (see Eq. (7.9) and Eq. (7.12) in
Ref.\cite{Ring80}). The limits of fully occupied ($u_i=0,v_i=1$) and
fully empty ($u_i=1,v_i=0$) levels have been carefully treated in
Refs. \cite{Avez12,Bertsch12,Robledo11} to avoid the collapse of the
overlap computation.

However, we note that in most realistic cases the $v_i$'s can be
extremely close to 0 or 1 but not exact 0 or 1. Strictly speaking,
these levels with such extreme emptiness or occupation should be
considered but may lead to exotic values ( extremely huge or
extremely tiny) of the Pfaffian in the proposed formulae. What is
worse, the Pfaffian values are easily out of the scope of the double
precision data type and cause the computation collapsed.

Careful treatment must be made to avoid such data overflow.
In this paper, we implement an accurate and
 reliable calculation for the overlap between arbitrary HFB vacua in a unified way.
 For the cases of ($u_i=1,v_i=0$) and ($u_i=0,v_i=1$), we treat them as the cases of ($u_i=1,v_i=\varepsilon$)
 and ($u_i=\varepsilon,v_i=1$), respectively. The tiny quantity $\varepsilon>0$ is chosen such that $\varepsilon^2$
 should be numerical zero relative to 1 in the practical calculation.
In other words, $\varepsilon$ should numerically satisfy
$1+\varepsilon^2=1$. Under this condition, $\varepsilon$ may be
chosen as large as possible so that the calculated Pfaffian values are not necessarily too huge or too tiny.
 For instance, one can choose $\varepsilon=10^{-8}$ when using double precision.
  Because $v_i^2(u_i^2)=\varepsilon^2$ is actually
 zero relative to $u_i^2(v_i^2)=1$ in practical calculations, this treatment does not change the HFB vacuum at all.
Therefore, without losing the generality, we assume that all levels in the Fock space are partly occupied,
 but some of their $u_i,v_i$ values are allowed to be extremely close to 0 or 1.
Ideally, $U, V$ are nonsingular in our assumption, and we can derive a new formula
 for the overlap between the HFB vacua based on the work of Bertsch and Robledo \cite{Bertsch12}.
 This formula is especially convenient for the symmetry restoration.
 Numerical calculations have been carried
out for heavy nuclear system to test the precision of the new formula by comparing with the Onishi formula.

In section \ref{theory}, the formalism of the new overlap formula is
given. Section \ref{calculation} provides an example of numerical
calculation. A summary is given in section \ref{summary}.

\section{The overlap between the HFB vacua}
\label{theory}

We denote $\hat c^\dagger_i$ and $\hat c_i$ as the creation and
annihilation operators defined in an $M$-dimensional Fock-space. The
Hartree-Fock-Bogoliubov(HFB) transformation is
\begin{equation}\label{hfb}
\left(\begin{array}{c} \hat\beta\\ \hat\beta^\dagger
\end{array}\right)
=\left(\begin{array}{cc}
U^\dagger&V^\dagger\\V^T&U^T
\end{array}\right)
\left(\begin{array}{c} \hat c\\\hat c^\dagger
\end{array}\right).
\end{equation}
Here, we assume $U$ and $V$ are nonsingular matrices, and their
shapes are $M\times M$. The HFB vacuum (unnormalized) can be written
as
\begin{eqnarray}\label{vacuum}
|\phi\rangle=\hat\beta_1\hat\beta_2...\hat\beta_M|-\rangle,
\end{eqnarray}
where $|-\rangle$ is the true vacuum.  By definition, one has
\begin{eqnarray}
\hat\beta_i|\phi\rangle=0\quad \mathrm{for}\quad 1\leq i \leq M.
\end{eqnarray}
The second HFB vacuum $|\phi'\rangle$ is defined in the same way,
but the prime, `$'$', is attached to the corresponding symbols to
show difference.

The overlap between $|\phi\rangle$ and$|\phi'\rangle$ is given by
\begin{eqnarray}
\langle\phi|\phi'\rangle&=&\langle -|\hat\beta^\dagger_M\hat\beta^\dagger_{M-1}...\hat\beta^\dagger_1\hat\beta'_1\hat\beta'_2...\hat\beta'_{M}|-\rangle\nonumber\\
&=&s_M\langle
-|\hat\beta^\dagger_1\hat\beta^\dagger_2...\hat\beta^\dagger_{M}\hat\beta'_1\hat\beta'_2...\hat\beta'_{M}|-\rangle,
\end{eqnarray}
where, $s_M=(-1)^{[M(M-1)/2]}$. If $M$ is even, $s_M=(-1)^{M/2}$.
Following the technique of  Bertsch and Robledo \cite{Bertsch12},
one can obtain
\begin{eqnarray}\label{overlap0}
\langle\phi|\phi'\rangle
&=&s_M\mathrm{pf}\left(\begin{array}{cc}
V^TU&V^TV'^*\\
-V'^\dagger V&U'^\dagger V'^*
\end{array}\right).
\end{eqnarray}
The shape of the  matrix in Eq. (\ref{overlap0}) is $2M\times 2M$, and no empty levels are omitted.
 For the norm overlap $\langle\phi|\phi\rangle$, it is real and positive. From Eq.(\ref{overlap0}) and the Bloch-Messiah theorem, one can get
\begin{eqnarray}\label{norm}
\langle\phi|\phi\rangle=s_M\mathrm{pf}\left(\begin{array}{cc}
V^TU&V^TV^*\\
-V^\dagger V&U^\dagger V^*
\end{array}\right)=\prod_{i=1}^{M/2}v_i^2.
\end{eqnarray}

 Denoting $\prod_{i=1}^{M/2}v_i$ by $\mathfrak{N}$,
the normalized quasi-particle vacuum, $|\psi\rangle$, can be
 written as
\begin{eqnarray}\label{psi}
 |\psi\rangle=\frac{|\phi\rangle}{\mathfrak{N}}.
\end{eqnarray}
Then, one finds that
\begin{eqnarray}\label{overlap1}
\langle\psi|\psi'\rangle=\frac{s_M}{\mathfrak{NN'}}\mathrm{pf}
\left(\begin{array}{cc}
V^TU&V^TV'^*\\
-V'^\dagger V&U'^\dagger V'^*
\end{array}\right).
\end{eqnarray}

In the symmetry restoration, the general rotational operator, involving
the spin and particle number projection, may be written as
\begin{eqnarray}
\hat {\mathbb{R}}(\Xi)=\hat R(\Omega)e^{-i \hat N \phi_n} e^{-i\hat
Z \phi_p},
\end{eqnarray}
where $\hat R(\Omega)$ is the rotation operator, and $\Omega$ refers
to the three Euler angles $\alpha,\beta,\gamma$. $e^{-i \hat N
\phi_n}$ and $e^{-i\hat Z \phi_p}$ are `gauge' rotational operators
induced by the  neutron and proton  number projection. $\hat N$ and
$\hat Z$ are neutron and proton number operators, respectively.
$\phi_n$ and $\phi_p$ are "gauge" angles for neutron and proton,
respectively. $\Xi$ refers to $(\Omega, \phi_n,\phi_p)$. The matrix
element $\langle\psi|\hat {\mathbb{R}}(\Xi)|\psi'\rangle$ needs to
be calculated. Let's define the general rotation transformation for
symmetry restoration,
\begin{eqnarray}
\hat {\mathbb{R}}(\Xi)\left(\begin{array}{cc} \hat c\\\hat c^\dagger
\end{array}\right)\hat {\mathbb{R}}^\dagger(\Xi)
=\left(\begin{array}{cc}
\mathbb{\mathbb{D}}^\dagger(\Xi)&0\\0&\mathbb{\mathbb{D}}^T(\Xi)
\end{array}\right)\left(\begin{array}{cc}
\hat c\\\hat c^\dagger
\end{array}\right),
\end{eqnarray}
where $\mathbb{D}_{ij}(\Xi)=\langle i|\hat {\mathbb{R}}(\Xi)|j\rangle$,
 and $|i(j)\rangle=\hat c^\dagger_{i(j)}|-\rangle$. The $\mathbb{D}(\Xi)$ matrix has
 the dimension $M\times M$.
%  From here on, we will skip the bracket
% $(\Xi)$ after the symbols $\hat {\mathbb{R}}$ and $\mathbb{D}$ to keep
% notation short.
 One can get
\begin{eqnarray}\label{rq}
\hat {\mathbb{R}}(\Xi)\left(\begin{array}{cc}
\hat\beta'\\\hat\beta'^\dagger
\end{array}\right)\hat {\mathbb{R}}^\dagger(\Xi)
=\mathcal{D}(\Xi) \left(\begin{array}{cc} \hat c\\\hat c^\dagger
\end{array}\right),
\end{eqnarray}
where
\begin{eqnarray}
\mathcal{D}(\Xi)=\left(\begin{array}{cc}
[\mathbb{D}(\Xi)U']^\dagger&[\mathbb{D}^*(\Xi)V']^\dagger \\
{[\mathbb{D}^*(\Xi)V']^T}&{[\mathbb{D}(\Xi)U']^T}
\end{array}\right).
\end{eqnarray}

By comparing Eq.(\ref{rq}) with Eq.(\ref{hfb}), one can obtain the
rotated overlap by replacing $U'$ and $V'$ in Eq.(\ref{overlap1})
with $\mathbb{D}(\Xi)U'$
 and $\mathbb{D}^*(\Xi)V'$, respectively. Thus
\begin{eqnarray}\label{overlap2}
\mathbb{N}_\mathrm{pf}(\Xi)=\langle\psi|\hat {\mathbb{R}}(\Xi)|
\psi'\rangle
=\frac{s_M}{\mathfrak{NN'}}\mathrm{pf}[\mathcal{M}(\Xi)],
\end{eqnarray}
where
\begin{equation}
\mathcal{M}(\Xi)=\left(\begin{array}{cc}
V^TU&V^T\mathbb{D}(\Xi)V'^*\\
-V'^\dagger \mathbb{D}^T(\Xi) V&U'^\dagger V'^*
\end{array}\right).
\end{equation}
This formula is essentially the same as the one proposed by Bertsch
and Robledo \cite{Bertsch12}, but we will transform it into a new
form. Supposing that there is a $\Xi_0$ satisfying
$\mathbb{N}_\mathrm{pf}(\Xi_0)\neq 0$, we have
\begin{eqnarray}\label{n1}
\frac{\mathbb{N}_\mathrm{pf}(\Xi)}{\mathbb{N}_\mathrm{pf}(\Xi_0)}
&=&\frac{\mathrm{pf}[\mathcal{M}(\Xi)]}{\mathrm{pf}[\mathcal{M}(\Xi_0)]}
=\frac{\mathrm{pf}\left[P\mathcal{M}(\Xi)P^T\right]}{\mathrm{pf}\left[P\mathcal{M}(\Xi_0)P^T\right]}\nonumber\\
&=&\frac{\mathrm{pf}[\mathcal{W}(\Xi)]}{\mathrm{pf}[\mathcal{W}(\Xi_0)]},
\end{eqnarray}
where
\begin{eqnarray}
\mathcal{W}(\Xi)=\left(\begin{array}{cc}
[U'V'^{-1}]^\dagger &-\mathbb{D}^T(\Xi)\\
\mathbb{D}(\Xi)&UV^{-1}
\end{array}\right),
\end{eqnarray}
and $P$ is
\begin{eqnarray}
P=\left(\begin{array}{cc}
0&(V'^\dagger)^{-1}\\
(V^T)^{-1}&0
\end{array}\right).
\end{eqnarray}
Therefore, one can get
\begin{eqnarray}\label{n3}
{\mathbb{N}_\mathrm{pf}(\Xi)}=\mathcal{C}{\mathrm{pf}[\mathcal{W}(\Xi)]},
\end{eqnarray}
where, the coefficient $\mathcal{C}$ is actually independent of
$\Xi_0$, and can be written as
\begin{eqnarray}
\mathcal{C}=\frac{\mathbb{N}_\mathrm{pf}(\Xi_0)}{\mathrm{pf}[\mathcal{W}(\Xi_0)]}
=\frac{s_M }{\mathfrak{NN'}\mathrm{det}P}={s_M\Delta\mathfrak{NN'}}.
\end{eqnarray}
Here, $\Delta$ is a phase determined by
\begin{eqnarray}
\Delta=\mathrm{det}D^*\mathrm{det}D'\mathrm{det}C\mathrm{det}C'^*.
\end{eqnarray}
In Eq.(\ref{n3}), we have used the Bloch-Messiah theorem and the
following equation
\begin{eqnarray}
\mathrm{det}P=\mathrm{det}[(V'^\dagger)^{-1}]\mathrm{det}[(V^T)^{-1}].
\end{eqnarray}

 Eq.(\ref{n3}) looks more convenient
 to be implemented and may save some computing time in contrast to Eq.(\ref{overlap2}),
 where extra evaluation of $V^T\mathbb{D}(\Xi)V'^*$ is required for each mesh point
 in the integral of projection.

For comparison, let us present a brief introduction of the overlap
of the Onishi formula \cite{Onishi66}. The unitary
transformation of the quasi-particles under rotation $\hat
{\mathbb{R}}(\Xi)$ can be written as
\begin{eqnarray}
\hat {\mathbb{R}}(\Xi) \left(\begin{array}{cc}
\hat\beta'\\\hat\beta'^\dagger
\end{array}\right)
\hat {\mathbb{R}}^\dagger(\Xi)=\left(\begin{array}{cc}
\mathbb{X}(\Xi)&\mathbb{Y}(\Xi)\\
\mathbb{Y}^*(\Xi)&\mathbb{X}^*(\Xi)\end{array}\right)\left(\begin{array}{cc}
\hat\beta \\ \hat\beta^\dagger
\end{array}\right),
\end{eqnarray}
where
\begin{eqnarray}
\mathbb{X}(\Xi)&=&U'^\dagger \mathbb{D}^\dagger(\Xi) U+V'^\dagger \mathbb{D}^T(\Xi) V \nonumber,\\
\mathbb{Y}(\Xi)&=&U'^\dagger \mathbb{D}^\dagger (\Xi)V^*+V'^\dagger
\mathbb{D}^T (\Xi)U^*.
\end{eqnarray}
The Onishi formula is then expressed as (see Ref.\cite{Schmid04}),
\begin{eqnarray}\label{onishi}
&&\mathbb{N}_\mathrm{Onishi}(\Xi)=\langle\psi|\hat
{\mathbb{R}}(\Xi)|
\psi'\rangle\nonumber\\
&=&(\pm)\sqrt{\mathrm{det}
[\mathbb{X}(\Xi)]}e^{-i(M_n\phi_n+M_p\phi_p)/2},
\end{eqnarray}
where $M_n$ and $M_p$ are the numbers of neutron and proton orbits
in the Fock space, respectively. The value of $\mathrm{det}
[\mathbb{X}(\Xi)]$ is a complex number, and the sign of the square
root is left undefined. Extra efforts must be made to determine the
sign before the application of the Onishi formula. For instance, in
the Projected Shell Model \cite{Hara95} without particle number
projection, the overlap between the BCS vacua is real and positive,
thus there is no sign ambiguity and the Onishi formula works.

\section{Numerical test of the overlap formulae}
\label{calculation}

 Although the sign problem is solved in Eq.(\ref{overlap2}) and Eq.(\ref{n3}),
 one can imagine that $\mathfrak{N}$, $\mathfrak{N}'$ are extremely tiny numbers by definition.
 Thus $\mathrm{pf}[\mathcal{M}(\Xi)]$ is also very tiny,
 but $\mathrm{pf}[\mathcal{W}(\Xi)]$ should be huge.
  Numerical accuracy of Eqs. (\ref{overlap2}) and (\ref{n3})
  needs to be carefully tested.
 It is believed that the Onishi formula is accurate except for its undetermined sign.
So, it is helpful to compare the numerical values of the overlaps
using
 Eq.(\ref{overlap2}), Eq.(\ref{n3}) and Eq.(\ref{onishi}).

To demonstrate the accuracy  and the reliability of the
 Eqs. (\ref{overlap2}) and (\ref{n3}), numerical calculations are performed
for the typical example of the deformed heavy nucleus $^{226}$Th.
For projection, we should take $|\psi\rangle=|\psi'\rangle$, and
then $\Delta=1$.

The $U,V$ matrices are obtained from the Nilsson+BCS method. The
single particle levels are generated from the Nilsson Hamiltonian
with the standard parameters \cite{Bengtsson85}. The single-particle
model space contains 5 neutron major shells with $N=$4-8 and 5
proton major shells with $N=$3-7, i.e., the Fock space has 145
neutron levels ($M_n=290$) and 110 proton levels ($M_p=220$). The
numbers of the active neutrons and protons are 96 and 70,
respectively. The quadrupole deformation is taken to be
$\epsilon_2=0.2$. Here, we only consider the axial symmetry for
simplicity.

In the no pairing case, the BCS vacuum becomes a pure slater
determinant, which is a challenge for Eq.(\ref{n3}) because all
$v_i$'s above the Fermi surface are zero. Consequently,
$\mathfrak{N}=0$ and $\mathcal{W}(\Xi)$ is meaningless due to the
singularity of $V$. Here, we use the double precision data type and
set $v_i=\varepsilon=10^{-8}$ for those $v_i=0$ orbits to avoid the
collapse of calculation. Therefore we have
\begin{eqnarray}
\langle\phi_n|\phi_n\rangle=(10^{-16})^{\frac{290-96}2}=10^{-1552},\nonumber\\
\langle\phi_p|\phi_p\rangle=(10^{-16})^{\frac{220-70}2}=10^{-1200},\nonumber
\end{eqnarray}
where, $|\phi_n\rangle$ and $|\phi_p\rangle$ are BCS vacua for
neutrons and protons, respectively, and
$|\phi\rangle=|\phi_n\rangle|\phi_p\rangle$. The tiny numbers
$10^{-1552}$ and $10^{-1200}$ are too far out of the scope of the
double precision data ($\sim10^{\pm307}$). To avoid the data
overflow, we multiply the tiny variable by $10^{200}$ several times
until the scaled absolute value falls into the interval
$[10^{-200},10^{200}]$. In other words, we use a number $y$ and an
integer number $k$ to express a tiny number $x$ through
$x=y\times(10^{-200})^k$. If $x$ is a huge number, then $k$ is
negative.

\begin{figure}
\centering
\includegraphics[width=3.0in]{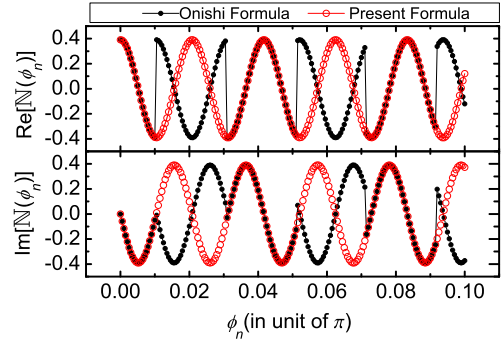}
\caption{(Color online) Overlaps of the ground state neutron slater
determinant for  $^{226}$Th as functions of $\phi_n$ with Euler
angles $\alpha=\gamma=0^\circ,\beta=10^\circ$, calculated with
present formula [Eq.(\ref{n3})] and the Onishi formula
[Eq.(\ref{onishi}) with `+' sign]. Re$[\mathbb{N}(\phi_n)]$ and
Im$[\mathbb{N}(\phi_n)]$ are the real and imaginary parts of the
overlap. } \label{ov1}
\end{figure}

However, for the Onishi formula of Eq.(\ref{onishi}), we do not need
to change $v_i=0$ to $v_i=\varepsilon$. The overlaps for the neutron
part, calculated with Eq.(\ref{n3}) and Eq.(\ref{onishi}),
 are compared in Fig.\ref{ov1}.  The curves of Eq.(\ref{n3}) are continuous, but
the sign uncertainty of Eq.(\ref{onishi}) causes the discontinuity.
However, if one copies the sign of Eq.(\ref{n3}) to
Eq.(\ref{onishi}),
 one can compare numerical difference between Eq.(\ref{n3}) and Eq.(\ref{onishi}) using
  the following quantity, $R$,
\begin{eqnarray}\label{r}
R=\left|\frac{\mathbb{N}_{\mathrm {Onishi}}(\phi_n)}{\mathbb{N}_{\mathrm {pf}}(\phi_n)}-1\right|.
\end{eqnarray}
 In all calculations, we found that $R<10^{-12}$ with double precision.
 This confirms that a small change of $v_i$ from zero to $\varepsilon$ almost does not
 affect numerical accuracy. However, it is crucial to keep Eq.(\ref{n3}) valid.
  Yet notice that $\mathbb{N}_{\mathrm {pf}}(\Xi)$ in Eq.(\ref{n3}) is obtained
   from a product of tiny and giant numbers.
    The same calculations have also been done with Eq.(\ref{overlap2}),
     and we also get $R<10^{-12}$. Thus we have presented an alternative way of
      using Eq.(\ref{overlap2}), where we set $v_i=\varepsilon$ for those empty orbits rather than omitting them\cite{Bertsch12}.

\begin{figure}
\centering
\includegraphics[width=3.0in]{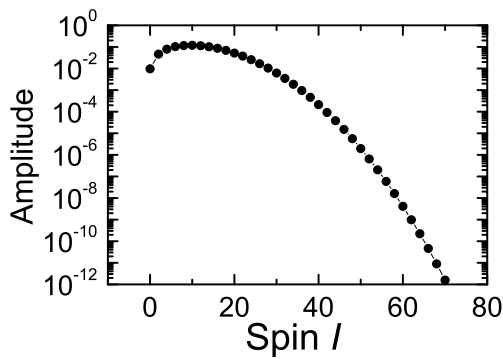}
\caption{ The amplitude of projection, $\langle\psi|\hat P^N\hat P^Z
\hat P^I_{00}|\psi\rangle$, as a function of spin $I$ at $N=96$ and
$Z=70$ using Eq.(\ref{n3}). $|\psi\rangle$ is the axially deformed BCS
vacuum but without pairing.} \label{am}
\end{figure}

Once the overlap is available, it is straightforward to perform the symmetry restoration.
 The deformed BCS vacuum of $^{226}$Th has been projected onto good particle number and spin.
  Therefore, one can test how precise the numerical calculations with Eq.(\ref{n3}) satisfy
\begin{eqnarray}\label{sum}
\sum_{N,Z,I}\langle\psi|\hat P^N\hat P^Z \hat
P^I_{00}|\psi\rangle=1,
\end{eqnarray}
where $\hat P^N$, $\hat P^Z$, and $\hat P^I_{MK}$ are
neutron-number, proton-number, and spin projection operators,
respectively. For the above vacuum state without pairing (i.e. the
ground state slater determinant), the particle numbers of both
neutrons and protons are good. Indeed, our particle number
projection (using 16 mesh points in the integral) shows that
$\langle\psi_n|\hat P^N|\psi_n\rangle=1$ ($N=96$), or 0 ($N\neq96$)
with numerical errors less than $10^{-13}$. Calculations for the
protons also have the same accuracy. This again shows the
reliability of Eq. (\ref{n3}). Angular momentum projection is also
performed on the same state in addition to the particle number
projection. The amplitude of $\langle\psi|\hat P^N\hat P^Z \hat
P^I_{00}|\psi\rangle$ with ($N=96, Z=70$) is plotted as a function
of spin $I$ in Fig.\ref{am}. In the integral of the spin projection,
100 mesh points are taken, and the range of spin is $0\leq I\leq70$,
and we indeed reproduced Eq.(\ref{sum}) with numerical error around
$10^{-12}$.

We also have tested Eq.(\ref{n3}) in the projection of the
triaxially deformed vacuum with normal pairing, which seems more
convenient to use Eq.(\ref{n3}). With the present method, similar
accuracy has also been achieved.

\section{Summary}
\label{summary}

 Following the strategy of Bertsch and Robledo \cite{Bertsch12},
we have proposed a new formula of the overlap between HFB vacua by
using the Pfaffian identity and assuming that the inverse of the $V$
matrix exists. This formula is especially convenient and efficient
in the symmetry restoration, and has the same high accuracy as the
Onishi formula as well as the correct sign. The reliability of the
present formula has been tested by carrying out the calculations of
the overlap and the quantum number projection for the heavy nucleus
$^{226}$Th. In the testing calculations, one has to be faced with two
numerical problems: (1) The extreme (huge or tiny) quantities are
certainly encountered, and we have properly treated this situation
to avoid data overflow (see the text). (2) For those empty orbits
with $v_i=0$, which make Eq.(\ref{n3}) invalid, one can change $v_i$
to a small quantity $\varepsilon(>0)$ to avoid the singularity
of $V$ matrix. It turns out that such treatments work very well.
Testing calculations have confirmed that the present formula is even
applicable to the pure slater determinant without losing the
numerical accuracy. Thus it is promising that Eq.(\ref{n3}) may be
applicable in evaluating the overlap between arbitrary HFB vacua.

\textbf{Acknowledgements} Z. G. thanks Prof. Y. Sun and Dr. F. Q.
Chen for the stimulating and fruitful discussions. The authors
acknowledge support from the National Natural Science Foundation of
China under Contract Nos. 11175258, 11021504 and 11275068.



%% References


%\bibliographystyle{model1a-num-names}



%\begin{thebibliography}{00}

\end{document}